\documentstyle[prl,aps,twocolumn,epsfig]{revtex}
\begin{document}
\draft

\title{Thermodynamics of Frustrated Quantum Spin Chains}

\author{K. Maisinger and U. Schollw\"{o}ck}
\address{Sektion Physik, Ludwig-Maximilians-Universit\"{a}t M\"{u}nchen,
Theresienstr.\ 37, 80333 Munich, Germany}

\date{March 9, 1998}

\maketitle
\begin{abstract}
We apply the transfer matrix DMRG to frustrated quantum spin chains, going down
to $T=0.025$ while being in the thermodynamic limit. The incommensurability 
problem of exact diagonalization and the negative sign problem of quantum Monte
Carlo vanish completely. To illustrate the method, we give results for chains 
with next-nearest neighbor frustration and the delta chain, which has been a 
testbed for many thermodynamic methods. By comparison, the DMRG proves to be 
an extremely powerful method for the old problem of the thermodynamics of 
frustrated systems.
\end{abstract}

\narrowtext

In the recent past, there has been enormous interest in one-dimensional
quantum systems, concentrating on the Hubbard and Heisenberg models.
As analytically exact methods such as the Bethe or the matrix product ansatz
can capture only some of the relevant models,
particular attention has been paid to the development of numerical methods.

Since one-dimensional strongly correlated systems show strong
quantum and thermal fluctuations, the most interesting physics is
to be expected close to or at $T=0$, where only quantum fluctuations
survive. To investigate such systems in the low temperature regime,
several numerical techniques are used, mainly exact diagonalizations,
quantum Monte Carlo, and the DMRG algorithm\cite{White 92}. 

In many cases, these methods yield completely satisfying, mutually
complementary or supporting results. Some systems, however, have
evaded thorough analysis so far. One of the strongest numerical challenges
comes from frustrated quantum spin models\cite{frust}. These models are
both of theoretical and experimental interest, which at the moment 
concentrates on the famous CuGeO$_{3}$\cite{germanium}.

\begin{figure}
\centering\epsfig{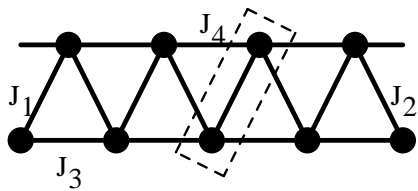}
\vspace{0.3truecm}
\caption{Hamiltonian studied by the transfer matrix DMRG. The dashed line
indicates spins grouped into a new object.}
\label{fig:ham}
\end{figure}

In this letter we show how the transfer matrix
DMRG can be applied to frustrated quantum spin systems, such that thermodynamic
quantities can be very reliably obtained down to $T\approx 0.01$ for
infinite system sizes, i.e.\ in the thermodynamic limit. 
There is no principal restriction of the systems that can be considered.
To discuss how this overcomes the difficulties of other methods,
we present numerical results for several physically relevant 
frustrated quantum spin systems. 

Let us first recall the other standard techniques.
Exact diagonalization techniques either calculate the full spectrum of a
very small system or extract information from the low-lying states of
somewhat longer systems\cite{Silver 94}. 
In both cases, finite size effects are present,
aggravated in frustrated systems by incommensurability effects. 

Both the quantum transfer matrix technique\cite{Betsuyaku 84}  
and quantum Monte Carlo start
from a Trotter decomposition which maps the one-dimensional quantum problem
to a two-dimensional classical problem. After a chequerboard
decomposition $H=H_1+H_2$\cite{Suzuki 76}, 
the partition function is given by
\begin{equation}
Z_{m} = \mbox{Tr\ } [e^{-\beta H_{1}/m}e^{-\beta H_{2}/m}]^{m},
\end{equation}
which becomes exact for the Trotter number $m\rightarrow\infty$.
In the quantum transfer matrix method, a transfer
matrix relating states on sites $i$ and $i+1$ is written down exactly and
thermodynamic quantities are 
calculated from its maximum eigenvalue. Matrix size grows
exponentially with $m$, restricting to a rather coarse Trotter 
decomposition, making access to low temperatures
difficult; however, results have no finite size effects. Quantum
Monte Carlo can access much larger Trotter numbers, while working on
finite system sizes. In frustrated spin systems, the essential limiting
factor comes from the well-known negative-sign problem: some of
the matrix elements of the infinitesimal evolution operator acquire negative
signs, but must be interpreted as positive probabilities. Only partial
elimination of this problem has been 
achieved\cite{Miyashita 95,Nakamura 95,Nakamura 97,Munehisa 94}.

Recently, the application of the DMRG to both classical\cite{Nishino 95}
and quantum\cite{Bursill 96} transfer matrices has been proposed. For the
Heisenberg chain, thermodynamic quantities can be calculated very 
well\cite{Wang 97,Shibata 97}; in fact, the results are also superior to those 
obtained from related attempts to apply DMRG techniques to 
thermodynamics\cite{Moukouri 96}. However, for the systems studied, other
methods had already yielded very good results\cite{Takahashi 71}.

Here, the transfer matrix DMRG is applied to the more challenging
frustrated spin-$\frac{1}{2}$ Hamiltonian
\begin{eqnarray}
{\cal H} &=& \sum_{i} J_1 {\mathbf S}_{2i} \cdot {\mathbf S}_{2i+1} +
J_2 {\mathbf S}_{2i-1} \cdot {\mathbf S}_{2i} + \nonumber \\
& & J_3 {\mathbf S}_{2i} \cdot {\mathbf S}_{2i+2} +
J_4 {\mathbf S}_{2i-1} \cdot {\mathbf S}_{2i+1} + HS^z_i ,
\end{eqnarray}
as shown in Figure \ref{fig:ham}. It comprises various frustrated spin models
of interest. 

The modifications to the original transfer matrix DMRG for an 
unfrustrated antiferromagnetic
chain\cite{Wang 97,Shibata 97} are very simple in principle. 
First, we rewrite the frustrated
spin chain as a spin ladder problem. If we group two neighboring
spins-$\frac{1}{2}$ into one object with 4 states, every second
nearest neighbor interaction is absorbed; the others and the next-nearest
neighbor interactions become effective nearest-neighbor interactions.
Similar in spirit to the restructuring method of Munehisa and 
Munehisa\cite{Munehisa 94}
and Nakamura's procedure for alleviating the negative sign problem in quantum
Monte Carlo\cite{Nakamura 97}, 
we are now free to choose any basis of the two-spin object 
which consists of eigenstates
of $S^{z}$, to maintain the good quantum number\cite{Nomura 91} of transfer 
matrix DMRG,
\begin{equation}
\sum_{j} (-1)^{j} (S^z_i)^j,
\end{equation}
where the sum runs along the Trotter direction.

For a $S=1/2$ problem, these
are essentially the Ising basis 
$|\uparrow\uparrow\rangle$, $|\uparrow\downarrow\rangle$, 
$|\downarrow\uparrow\rangle$, $|\downarrow\downarrow\rangle$ 
and the singlet-triplet
basis $|\uparrow\uparrow\rangle$, $(|\uparrow\downarrow\rangle +
|\downarrow\uparrow\rangle)/\sqrt{2}$, 
$(|\uparrow\downarrow\rangle -|\downarrow\uparrow\rangle)/\sqrt{2}$, 
$|\downarrow\downarrow\rangle$. Numerically,
we find that both yield the same results, though the latter has no update
of its internal bond in a transfer matrix step and might therefore be
expected to be more precise. All results given below were obtained with the
singlet-triplet basis.

The transfer matrix DMRG is governed by the parameters $M$, the
number of states kept, and $\beta_0$, the initial temperature and at the
same time the stepwidth of the values of $\beta$ considered during the
DMRG iterations. As the two-spin object has 4 instead of 2 states, for given
computational resources, $M$ has to be halved. To estimate the effect on
DMRG precision (and also check the correctness of the program),
we applied both the original and our transfer matrix DMRG to the unfrustrated
Heisenberg chain. We find that the new version shows much better results
for $M/2$ than the original one for $M$ states kept, as every second bond
is taken into account exactly. This greatly enhances the applicability of this
method also to frustrated chains. To give one result, the internal
energy per site of the isotropic Heisenberg model should go to $-0.4431$ for
$T\rightarrow 0$. For $M=32$, the original transfer matrix 
DMRG extrapolates $-0.448$ for 
$\beta_0=0.2$, ours gives $-0.445$ for $M=16$ only.

\begin{figure}
\centering\epsfig{file=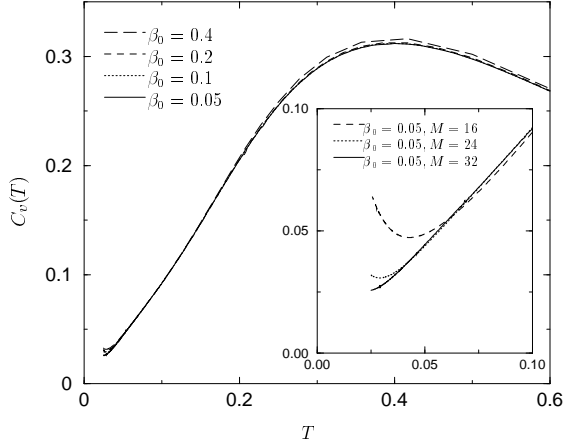,scale=0.5}
\vspace{0.7truecm}
\caption{Specific heat of the next-nearest neighbor frustrated
spin-$\frac{1}{2}$ chain at its phase transition ($M=32$). The inset
shows the convergence with the number $M$ of states kept 
of the $\beta_0=0.05$ curve for very low
temperatures.}
\label{fig:frust24cv}
\end{figure}

There are several advantages of the transfer matrix DMRG for frustrated 
spin systems. As opposed to the exact 
diagonalisation of small systems, we have no incommensurability 
problem, as we work in the thermodynamic limit right away. Unlike 
quantum Monte Carlo, there is no such thing as a negative sign problem;
also quantum MC results are obtained in finite systems, 
albeit often rather large
ones. Compared to the recent progress made by Nakamura\cite{Nakamura 97}, 
DMRG works for 
{\em all} frustrated spin systems; furthermore, there is no problem with
the susceptibility. Quantum transfer matrix methods are closest in
spirit to the transfer matrix DMRG. They relate to each other exactly
as $T=0$ DMRG relates to exact diagonalisation. Like $T=0$ DMRG that can treat
much longer systems than exact diagonalization, 
here much higher Trotter numbers can be obtained:
Typically, in quantum transfer matrix problems
one is restricted to less than 10 Trotter slices when dealing with frustrated 
problems. The transfer matrix DMRG has no fixed number of Trotter slices; 
it increases with $\beta$. Consider
$T=0.025$, or $\beta=40$ as a typical low temperature one might wish
to reach. For a realistic starting temperature $\beta_{0}=0.05$, this implies
$800$ Trotter slices
at that temperature, or 800 DMRG steps. While maintaining the
advantages of the quantum transfer matrix, the problem of convergence
of results in the Trotter number can be virtually eliminated, as the results
presented here show. However, the problem of convergence in the number $M$ of states
kept has also to be considered. Working with $M\leq 32$, which is still 
possible on a workstation, we find that down to $T\approx 0.05$ results
are very well converged; for lower $T$ the trend is very clear (see as an 
example
the inset of Figure \ref{fig:frust24cv}).  

In order to test the transfer matrix DMRG, we have first applied it to the
so-called delta chain,
because it exhibits a rather special low-temperature behavior\cite{Kubo 93},
of which some approximative analytic results are known\cite{Kubo 96}. 
Numerical methods\cite{Miyashita 95,Nakamura 95,Kubo 93}
have been applied to this chain with varying success, which makes this
a good starting point for a comparison of results.

The delta chain consists of spin triangles, which are linked through the
triangle basis spins ($J_1=J_2=J_3=1$, $J_4=0$). 
It can be understood as a frustrated chain with
next-nearest neighbor interactions, of which every second has been taken
out. It has the same doubly degenerate ground state as the 
Majumdar-Ghosh model, with spin singlets sitting either on all left or on all
right sides of the triangles\cite{exact}. 
Numerical studies reveal that the delta chain gap
$\Delta\approx 0.22$ is much smaller than the Majumdar-Ghosh
model\cite{Majumdar 69} gap, which suggests a rather delocalized 
excitation\cite{Kubo 93}.
On the other hand, exact diagonalisation shows an almost flat dispersion
relation, indicating a localized excitation; these results are however
strongly size-dependent\cite{Kubo 93}. It has been suggested 
that there
are two types of excitations in the chain, kinks and 
antikinks\cite{Kubo 96,Sen 96}. 
Elementary excitations can be imagined as domain walls between ground state
segments
with a singlet on the left to singlet on the right or right to left flip. 
In the former case, it is
energetically favorable to insert one triangle with a singlet on the
basis. This is low in energy, being an eigenstate of the local triangle
Hamiltonian, but localized. In the latter case, one triangle has no singlet
at all, which is high in energy. However, this triangle can be
delocalized. Obviously, kink and antikink excitations must alternate.
Thus the spectrum finds a natural explanation. 

For the specific heat, the presence of two different excitation types
with different energies implies a two-peak structure, with a conventional
high-temperature peak and an additional low-temperature peak. 

We have calculated the specific heat $C_{v}(T)$ by numerical differentiation
of the internal energy which was directly evaluated by the DMRG, starting
from temperatures $\beta_{0}=0.5$, $0.4$, $0.2$, $0.1$ and $0.05$, while
retaining $M=16$, $24$ and $32$ states, down to temperature $\beta=40$.
This implies maximum Trotter numbers $m=80$, $100$, $200$, $400$ and $800$.

\begin{figure}
\centering\epsfig{file=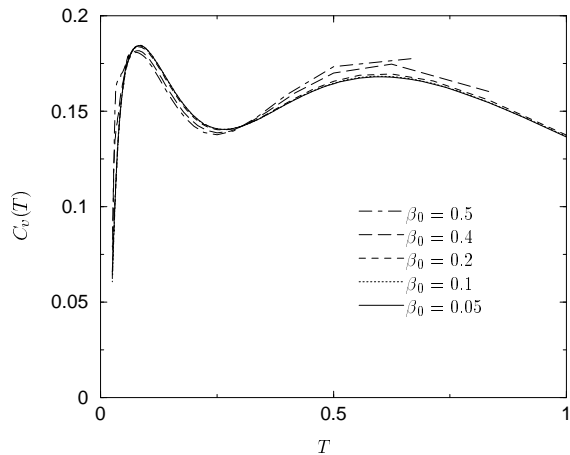,scale=0.5}
\vspace{0.3truecm}
\caption{Specific heat of the delta chain for 32 states kept and various
$\beta_0$. Note the different convergence behavior of the high- and the low-$T$
peak.}
\label{fig:deltacv1}
\end{figure}

\begin{figure}
\centering\epsfig{file=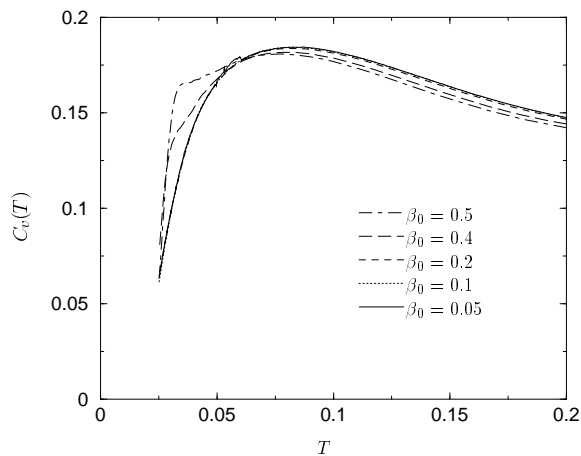,scale=0.5}
\vspace{0.3truecm}
\caption{Low temperature peak of the specific heat of the delta chain
($M=32$). Curves for
$\beta_0=0.1$ and $0.05$ are almost indistinguishable.}
\label{fig:deltacv2}
\end{figure}

We find that the low-temperature peak in the specific heat is converged
both in $\beta_{0}$ and $M$; 
we extract thermodynamic
limit values of $T_{peak}=0.082(2)$ for the peak temperature and a value
$C_{v}^{max}=0.184(1)$. Calculating
the specific heat from a variational ansatz, Nakamura and Kubo\cite{Kubo 96} 
predict
this second peak at $T\approx 0.05$, with a height of $C_v\approx 0.24$. 
Numerically, Kubo finds $C_{v}^{max}\approx 0.18$ at $T_{peak}\approx 0.12$
from a quantum transfer method\cite{Kubo 93}; 
these are unextrapolated results from his
maximum Trotter number. Otsuka finds 
$C_{v}^{max}\approx 0.20$ at $T_{peak}\approx 0.12$ from a
recursion method\cite{Otsuka 95}. 
Monte Carlo studies have reached the 
high-temperature slope of the low-$T$ peak of $C_{v}$, but could not localize
and estimate it reliably\cite{Nakamura 95}. 
The DMRG therefore clearly gives the quantitatively most
accurate results; Nakamura's modified Monte Carlo\cite{Nakamura 97} 
should in principle
be capable of a similar precision for the problem under study; it must
be stressed that this is because the delta chain fits his negative sign
problem elimination formula which is however not universal. The DMRG exploits
no special feature of the studied frustrated system and is thus more
versatile. 

Let us comment on the convergence of the specific heat with 
$\beta_0\rightarrow 0$. The underlying internal energy is uniformly 
underestimated
for finite $\beta_0$, converging as $\beta_0^2$. However, 
as can be seen in Figure \ref{fig:deltacv1},
for coarse Trotter slicing, the high-temperature peak is overestimated,
while the low-temperature peak is underestimated (a similar phenomenon is 
also visible in Ref.\ \onlinecite{Kubo 93}).
This can be understood in terms
of a ``sum rule''. Obviously,
\begin{equation}
s(T=\infty)-s(T=0) = \int_{0}^{\infty} \frac{C_{v}(T)}{T} dT.
\end{equation}
Now all the classical models resulting from the Trotter decomposition have
vanishing zero-temperature entropy, while they all map free spins for
$T\rightarrow\infty$, i.e.\ $s(T=\infty)=\ln 2$, yielding a
constant value of the integral for all $\beta_0$. Thus the errors due to
the Trotter decomposition have to
compensate each other; a coarse decomposition will miss low-$T$ features
and overemphasise high-$T$ features. The shoulders for $\beta=0.5$ and $0.4$
in Fig.\ \ref{fig:deltacv2} we interpret as a low-$T$ cutoff of $C_v$ due
to large $\beta_0$.

For a comparison, the specific heat (Fig.\ \ref{fig:frust24cv}) of 
a typical next-nearest neighbor 
frustrated chain (the spin-$\frac{1}{2}$ chain at
its transition point $J_1=J_2=1$, $J_{3}=J_{4}=0.2411$) 
has a one-peak structure, which
is converged down to $T\approx 0.04$ and can be linearly extrapolated
to $C_{v}=0$ at $T=0$. For very low temperatures, the results deviate towards
higher values. We find that both a too coarse
Trotter decomposition and a too small $M$ lead to 
an overestimation of $C_{v}$; 
comparing the
curves for $\beta_{0}=0.1$ and $\beta_{0}=0.05$ indicates that the remaining
error is almost entirely due to the finite number $M$ of DMRG
states. As the inset in Fig.\ \ref{fig:frust24cv} shows, even for very
low $T$ the results converge fast with $M$.

\begin{figure}
\centering\epsfig{file=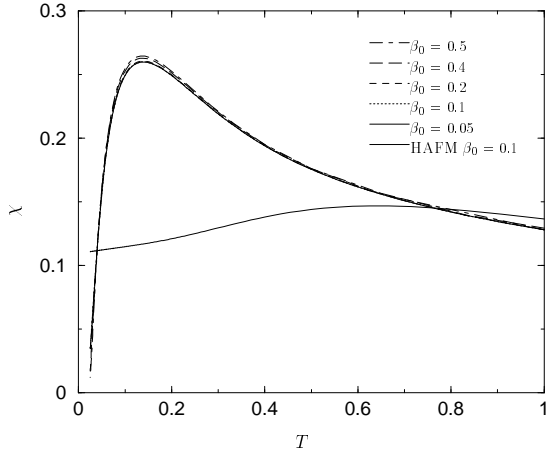,scale=0.5}
\vspace{0.3truecm}
\caption{Susceptibility of the delta chain ($M=32$) compared
to the susceptibility of the Heisenberg chain ($M=32$).}
\label{fig:susc}
\end{figure}

Let us finish by showing the susceptibility of the delta chain in comparison
to the exactly known susceptibility of the isotropic $S=\frac{1}{2}$ 
Heisenberg chain (Fig.\ \ref{fig:susc}), both calculated by numerical
differentiation of the explicitly evaluated magnetisation.
Whereas for the latter one finds a finite $\chi$ for $T=0$, here
the susceptibility is exponentially activated due to the gap in the
excitation spectrum. No particular physics is hidden there, but from the
point of view of the transfer matrix DMRG it can be observed that again
results converge very fast in $\beta_0$ and $M$.

To conclude, we have demonstrated how the transfer matrix DMRG overcomes the
problem of obtaining the low-temperature thermodynamics of frustrated spin
chains, while at the same time working in the thermodynamic limit without
serious restrictions on the Hamiltonians considered. The method can easily be
upgraded to more complicated frustrated chains, where other ad hoc fixtures
fail. We are therefore confident that the transfer matrix DMRG will emerge
as the most convenient method for obtaining the thermodynamics of frustrated
spin or also frustrated fermionic systems.

All calculations were carried out on a 533 Mhz alpha with Linux and a 16
processor T3E.


\begin{references}
\bibitem{White 92} S.R. White, Phys.\ Rev.\ Lett.\ {\bf 69}, 2863 (1992);
Phys.\ Rev.\ {\bf B 48}, 10345 (1993).
\bibitem{frust} see, for examples, H.T. Diep (ed.), {\em Magnetic
Systems with Competing Interactions (Frustrated Spin Systems)}, World
Scientific, Singapore, 1994.      
\bibitem{germanium} M. Hase, I. Terasaki, K. Uchinokura, 
Phys.\ Rev.\ Lett.\ {\bf 70}, 3651 (1993); 
G. Castilla, S. Chakravarty, and V. J. Emery,
Phys.\ Rev.\ Lett.\ {\bf 75}, 1823 (1995).    
\bibitem{Silver 94} R.N. Silver and H. R\"oder, Int. J. of Modern Physics,
{\bf C 5}, 735 (1994).
\bibitem{Betsuyaku 84} H. Betsuyaku, Phys.\ Rev.\ Lett.\ {\bf 53}, 629 (1984);
Prog.\ Theor.\ Phys.\ {\bf 73}, 319 (1985).
\bibitem{Suzuki 76} M. Suzuki, Prog.\ Theor.\ Phys.\ {\bf 56}, 1454 (1976).
\bibitem{Miyashita 95} S. Miyashita, J. Phys.\ Soc.\ Jpn.\ {\bf 63}, 2449 (1994).
\bibitem{Nakamura 95} T. Nakamura and Y. Saika, J. Phys.\ Soc.\ Jpn.\ {\bf 64},
695 (1995). 
\bibitem{Nakamura 97} T. Nakamura, preprint cond-mat 9707019.
\bibitem{Munehisa 94} T. Munehisa and Y. Munehisa, Phys.\ Rev.\ {\bf B 49}, 
3347 (1994). 
\bibitem{Nishino 95} T. Nishino, J. Phys.\ Soc.\ Jpn.\ {\bf 64}, L3598 (1995).
\bibitem{Bursill 96} R.J. Bursill, T. Xiang, and G.A. Gehring, J. Phys.\
Cond.\ Matt.\ {\bf 8}, L583 (1996).
\bibitem{Wang 97} X. Wang and T. Xiang, Phys.\ Rev.\ {\bf B 56}, 5061 (1997).
\bibitem{Shibata 97} N. Shibata, J. Phys.\ Soc.\ Jpn.\ {\bf 66}, 2221 (1997).
\bibitem{Moukouri 96} S. Moukouri and L.G. Caron, Phys.\ Rev.\ Lett.\
{\bf 77}, 4640 (1996).
\bibitem{Takahashi 71} M. Takahashi, Prog.\ Theor.\ Phys.\ {\bf 46}, 401
(1971), Prog.\ Theor.\ Phys.\ {\bf 51}, 1348 (1974); 
M. Gaudin, Phys.\ Rev.\ Lett.\ {\bf 26}, 1301 (1971); S. Eggert,
I. Affleck, and M. Takahashi, Phys.\ Rev.\ Lett.\ {\bf 73}, 332 (1994).
\bibitem{Nomura 91} K. Nomura and M. Yamada, Phys.\ Rev.\ {\bf B 43}, 8217
(1991).
\bibitem{exact} T. Hamada {\em et al.},
J. Phys.\ Soc.\ Jpn.\ {\bf 57}, 1891 (1988); B. Doucot and I. Kanter,
Phys.\ Rev.\ {\bf B 39}, 12399 (1989); F. Monti and A. S\"{u}t\"{o},
Phys.\ Lett.\ {\bf A 156}, 197 (1991); Helv.\ Phys.\ Acta {\bf 65}, 560
(1992).
\bibitem{Majumdar 69} C.K. Majumdar and D. Ghosh, J. Math.\ Phys.\ {\bf 10},
1388 (1969); C.K. Majumdar, J. Phys.\ C {\bf 3}, 911 (1970).
\bibitem{Kubo 93} K. Kubo, Phys.\ Rev.\ {\bf B 48}, 10552 (1993). 
\bibitem{Kubo 96} T. Nakamura and K. Kubo, Phys.\ Rev.\ {\bf B 53}, 6393 (1996).
\bibitem{Sen 96} D. Sen {\em et al.}, Phys.\
Rev.\ {\bf B 53}, 6401 (1996).
\bibitem{Otsuka 95} H. Otsuka, Phys.\ Rev.\ {\bf B 51}, 305 (1995).
\end{references}
\end{document}